\documentclass[preprint,aps,prc,showpacs,nofootinbib]{revtex4}
\usepackage{epsfig}
\usepackage{graphicx,amsmath}

\newcommand{\Li}{\mbox{Li}_2}

\newcommand{\dd}{\mbox{d}}
\newcommand\ba{\begin{eqnarray}}
\newcommand\ea{\end{eqnarray}}
\newcommand\nn{\nonumber}
\newcommand{\be}{\begin{equation}}
\newcommand{\ee}{\end{equation}}
\def\Li#1#2{{\mathrm{Li}}_{#1}\left(#2\right)}

\begin{document}
\title{Charge asymmetry for electron (positron)-proton elastic scattering at large angle}

\author{E. A. Kuraev, V. V. Bytev, S.~Bakmaev}
\affiliation{\it JINR-BLTP, 141980 Dubna, Moscow region, Russian
Federation}
\author{E.~Tomasi-Gustafsson}
\affiliation{\it DAPNIA/SPhN, CEA/Saclay, 91191 Gif-sur-Yvette
Cedex, France }

\date{\today}

\begin{abstract}
Charge asymmetry in electron (positron) scattering arises from the interference of the Born amplitude and the box-type amplitude corresponding to two virtual photons exchange. It can be extracted from electron proton and positron
proton scattering experiments, in the same kinematical conditions. Considering the virtual photon Compton scattering tensor, which contributes to the box-type amplitude,
we separate proton and inelastic contributions in the intermediate state and parametrize the proton form-factors as the sum of a pure QED term and a strong interaction term. Arguments based on analyticity are given in favor of cancellation of contributions from proton strong interaction form factors and  of inelastic intermediate states in the box type amplitudes.  In frame of this model, with a realistic expression for nucleon form-factors, numerical estimations are given for moderately high energies.
\end{abstract}

\maketitle
\section{Introduction}

The nucleon structure is traditionally investigated using electromagnetic probes, and assuming that the interaction occurs through the exchange of a virtual photon, which carries the momentum transfer, $q$,  from the incident to the scattered lepton. Recently, large attention was devoted to $2\gamma$ exchange amplitude both in scattering and annihilation channels \cite{Re04,china,twogamma,By07}, in connection with experimental data on electromagnetic proton form factors (FFs) \cite{Jo00}.

The extraction of box type (two photon exchange amplitude (TPE)) contribution to the elastic
electron-proton scattering amplitude
is one of long standing problems of experimental physics. It can be obtained from electron proton and positron proton scattering at the same kinematical conditions. A similar information about
TPE amplitude in the annihilation channel can be obtained from the measurement of the forward-backward asymmetry in  proton-antiproton production in  electron-positron annihilation (and from the time-reversal process).

The theoretical description of TPE amplitude is strongly model dependent. Two reasons should be mentioned: the experimental knowledge of nucleon FFs is restricted in a small kinematical region,  and the contribution of the intermediate hadronic states can be only calculated with large uncertainty,
the precision of the data being insufficient to constrain the models.

A general approximation for proton electromagnetic form-factors follows the  dipole approximation:
\be
G_E(q^2)=\displaystyle\frac{G_M(q^2)}{\mu} =G_D(Q^2)=(1+Q^2/0.71\mbox{~GeV}^2)^{-2},Q^2=-q^2=-t,
\ee
where $\mu$ is the anomalous magnetic moment of proton.
However, recent experiments \cite{Jo00} showed a deviation of the proton electric FF from this prescription, when measured following the recoil polarization method \cite{Re68}, which is more precise than the traditional Rosenbluth separation \cite{Ro50}. Such deviation was tentatively explained, advocating the presence of two photon exchange contribution.

The motivation of this paper is to perform the calculation of charge odd correlation
\begin{gather}
A^{odd}=\frac{\dd\sigma^{e^-p}-\dd\sigma^{e^+p}}{2d\sigma_B^{ep}},
\label{eq:asym}
\end{gather}
in the process of electron-proton scattering in frame of an analytical model (AM), free from uncertainties connected with inelastic hadronic state in intermediate state of the TPE amplitude. In frame of this model it is possible to show that there is a compensation between the effects due to  strong interaction FFs and those due to the inelastic intermediate states, within an accuracy discussed below.

Our paper is organized as follows. In part II,
we introduce a new decomposition of proton FFs, separating QED and strong interaction contributions. In section III we formulate the analytical model and calculate the contribution of the QED part of FFs.  In Section IV the resulting expression for the asymmetry is obtained. In Section V we present the results of numerical integration for asymmetries
and in the Conclusions we estimate the accuracy of the obtained results. The Appendix contains some details of the calculation.

\section{New form of proton form factors}

The cross-section of elastic ep scattering
\begin{gather}
e(p_1)+p(p)\to e(p_1')+p(p')
\end{gather}
in Born approximation, in laboratory (Lab) frame ($p=(M,0,0,0)$,  $p_1=E(1,1,0,0))$ has the form:
\begin{gather}
\frac{\dd\sigma_B}{\dd\Omega}=\frac{\sigma_M\sigma_{red}}{\varepsilon(1+\tau)},
\quad
\sigma_M=\frac{\alpha^2\cos^2\frac{\theta}{2}}{4E^2\sin^4\frac{\theta}{2}}\frac{1}{\rho},
\quad
\rho=1+\frac{2E}{M}\sin^2\frac{\theta}{2},
\quad \tau=\frac{Q^2}{4M^2},
\\ \nonumber
t=\frac{s(1-\rho)}{\rho},~ Q^2=-q^2=-t=2p_1p_1',\quad s=2ME,\quad u=-\frac{s}{\rho}=-2p p_1',
\\ \nonumber
s+t+u=0,\quad \varepsilon^{-1}=1+2(1+\tau)\tan^2\frac{\theta}{2},
\end{gather}
with
\begin{gather}
\sigma_{red}=\tau G_M^2+\varepsilon G_E^2,\quad
 G_M=F_1+F_2, \quad G_E=F_1-\tau F_2.
\end{gather}
Here $\theta$- is the Lab electron scattering angle.

In the analysis of the TPE amplitude we consider the electromagnetic interactions in the lowest order of
perturbation theory. Hadron electromagnetic FFs in the space-like region, which  parametrize the interaction
with the external electromagnetic vertex as
\ba
\bar{u}(p')[\Gamma_1(q^2)\gamma_\mu+\frac{1}{2M}\hat{q}\gamma_\mu\Gamma_2(q^2)] u(p), q=p'-p, p^2=p^{'2}=M^2,
\ea
are functions of one kinematical variable, $q^2$. The
static value of the Dirac FF of the proton (for $Q^2$=0) is unity due to 'QED'  origin, i.e. interaction with a point-like quark. Therefore we decompose the Dirac FF in two terms, corresponding to different effective internal momenta:
\be
F_1(q^2)=F_{1Q}(q^2)+F_{1s}(q^2),\quad  F_2(q^2)=F_{2s}(q^2),\mbox{~with~} F_{1s}(0)=0,\mbox{~and~} F_{2s}(0)=\mu ,
\label{eq:eq3}
\ee
where the subscript $'s'$ stands for strong interaction and $F_{1Q}$ corresponds to the 'QED' contribution. $F_{1Q}(q^2)$ is relevant at very small momenta compared to the typical hadronic mass and describes the exchange of gluons with virtuality $m_q^2\ll k_g^2\ll Q^2$, where $m_q$ is the mass of current quarks. $F_{1Q}$ is a fast decreasing function of $Q^2$, and may be represented by an expression of Sudakov type FF:
\cite{KF78}:
\be
F_{1Q}(q^2)=exp\left ({-\frac{\alpha_s C_F}{4\pi}\ln^2\frac{Q^2}{m_q^2}}\right ).
\label{eq:Sudakov}
\ee
Contrary to the charged lepton Dirac form-factor,
which can be considered constant and equal to unity, independently on $Q^2$,
the Sudakov parametrization gives a negligible contribution as $Q^2$ deviates from zero. Although the Sudakov parametrization can not, in principle, be applied for $Q^2\gg m_q^2$, taking $m_q\sim$ 3-5 MeV, the parametrization from (\ref{eq:Sudakov}) applies starting from the quark mass value. This justifies the following parametrization to the hadronic electromagnetic FFs that will be used later on for practical purposes:
\ba
F_{1Q}(Q^2)=1, Q^2<m_0^2\sim m_\pi^2; \nn \\
F_{1Q}(Q^2)=0, Q^2>m_0^2.
\label{eq:eqsu}
\ea
In most of the kinematical domain, one can safely neglect the interference between $F_{1Q}$ and $F_{1s}$, as they act in different regions of transferred momenta: $F_1^2= F_{1Q}^2+F_{1s}^2$. However, the derivative at $Q^2\to 0$, which is connected to the charge radius, $r_p$, is sensitive to the interference term:
$$\frac{d}{dQ^2}\left . F_1^2(Q^2)\right |_{Q^2=0}=2F_{1s}'(0).$$
Parametrization of space-like form factors in terms of fractional polynomials have already been suggested in literature \cite{Brash,Kelly}. The present choice of the strong interaction FF is consistent with the following parametrization:
\be
 F_{1s}(Q^2)=\displaystyle\frac{Q^2 r_p^2 \left [1+ \sum_1^n c_k
 \left (\displaystyle\frac{Q^{2}}{Q_0^{2}}\right )^k\right ]}
 {6 \left [ 1+ \sum_1^{n+3} d_k
 \left (\displaystyle\frac{Q^{2}}{Q_0^{2}}\right )^k\right ]},~
 F_{2s}(Q^2)=\displaystyle\frac{\mu  \left[1+ \sum_1^n e_k
 \left (\displaystyle\frac{Q^{2}}{Q_0^{2}}\right )^k\right ]}
 {  1+ \sum_1^{n+3} f_k
 \left (\displaystyle\frac{Q^{2}}{Q_0^{2}}\right )^k }.
\label{eq:eqff}
\ee
where $c$, $d$, $e$,$f$ can be considered as fitting parameters. The last coefficients of the series $c_n$, $d_{n+3}$, $e_n$, $f_{n+3}$ are constrained by the high $Q^2$ asymptotic limit. The low $Q^2$ properties: $F_{1s}(O)=0$, $F'_{1s}(0)= \displaystyle\frac{1}{6}r_p^2$ and $F_{2s}(0)=\mu$, where $\mu$ is the magnetic moment, are explicitly taken into account.

\section{Formulation of the analytical model}

Let us discuss now the arguments in favor of mutual cancellation of the terms of order of $F_s^2$ with the
contribution of the inelastic hadronic intermediate states, in TPE amplitude.

The TPE amplitude contains the virtual photon Compton scattering tensor.
It can be split in two terms, when only strong
interaction contributions to Compton amplitude are taken into account. One term
(the elastic term) is the generalization of the Born term with the strong-interaction FFs at the vertexes of
the interaction of the virtual photons with the hadron. We suppose that the hadron before and after the
interaction with the photons remains unchanged. The second term (inelastic) corresponds to inelastic channels
formed by pions and
nucleons or the excited states of the nucleon such as the $\Delta$ resonance.

\begin{figure}
\begin{center}
\includegraphics[width=12cm]{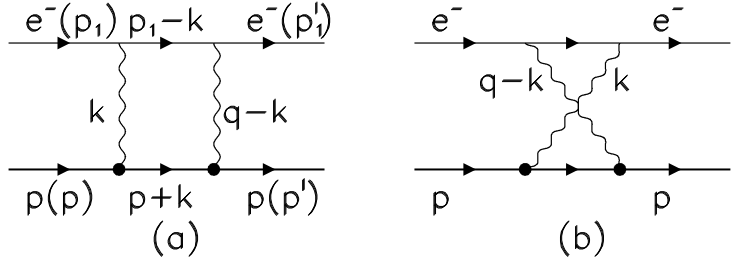}
\caption{\label{Fig:twogam} Feynman diagrams for two-photon
exchange in elastic $ep$ scattering: box diagram (a)
and crossed box diagram (b).}
\end{center}
\end{figure}

\begin{figure}
\begin{center}
\includegraphics[width=12cm]{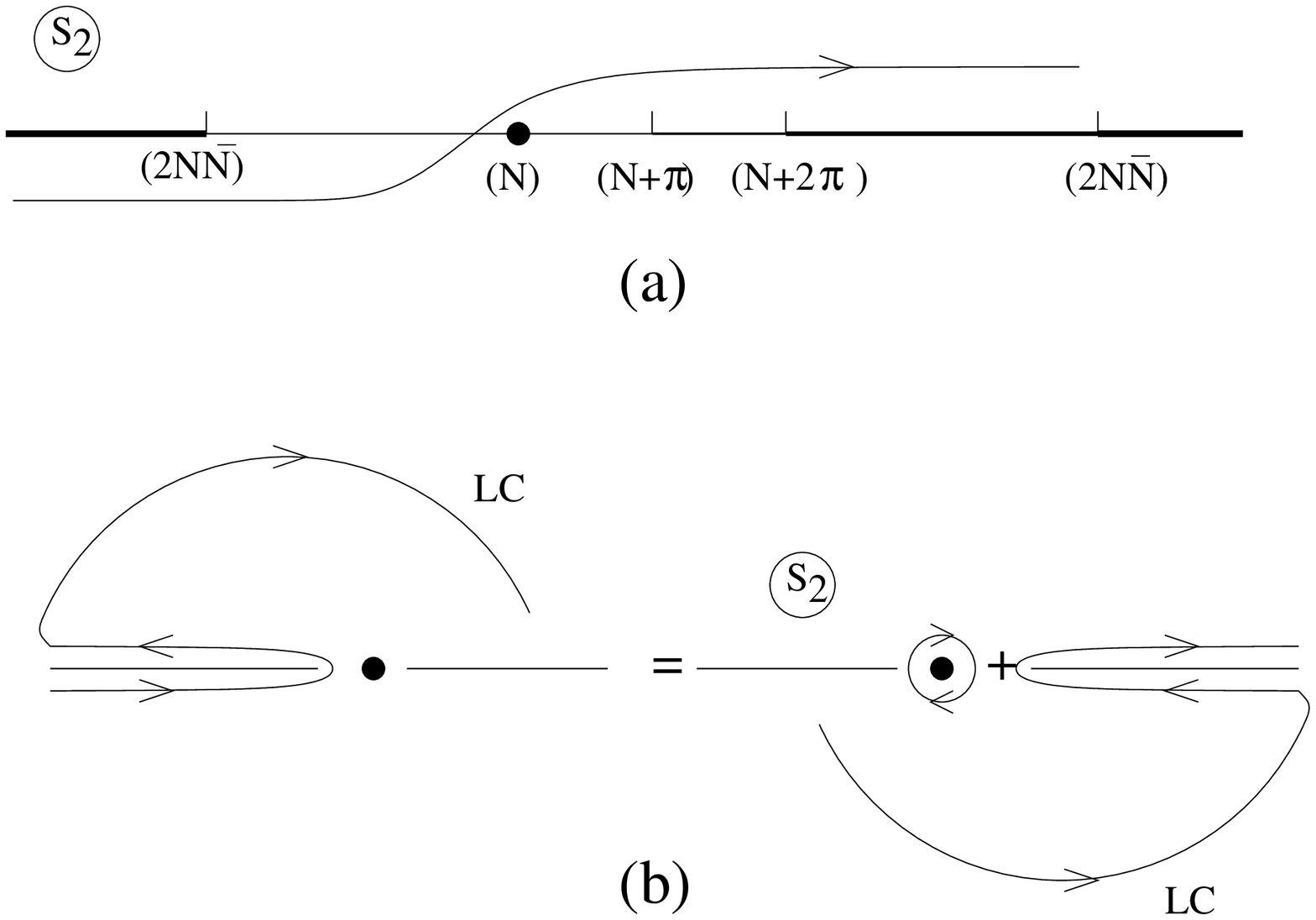}
\caption{\label{Fig:contour} Illustration of singularities along the $s_2$ real axis with the open contour C (a), and with the contour C closed (b). LC stays for large circle contribution.}
\end{center}
\end{figure}

Denoting the loop momenta in TPE amplitude as $k$, the loop momenta phase volume can be written as:
$$d^4k=\frac{1}{2s} d^2k_{\perp}ds_e ds_p,$$
where $s_e=(p_--k)^2$ and $s_p= (p+k)^2$ are the invariant mass squared of the electron and proton blocks, and $ k_{\perp}$ is a two-dimensional euclidean vector orthogonal to the momenta of the initial particles :$ k_{\perp}p_-=
 k_{\perp}p=0$. Considering both Feynman diagrams, the box and the crossed diagram (Fig. \ref{Fig:twogam}), the integration over $s_e$ ($-\infty<s_e<\infty $) can be done calculating the residue by $s_e$ in the electron Green function (omitting higher QED corrections).
The contour for the $s_p$ integration has a Feynman contour form: $-\infty-i0<s_p<\infty+i0$ (see Fig. \ref{Fig:contour}).

Let now consider the analytic properties of the Compton scattering amplitude with both photons off-mass shell as a function of the complex variable $s_p$.
The singularities on the physical sheet are a pole, located at $s_p=M^2$, which corresponds to a proton in the intermediate state and a series of cuts corresponding to inelastic states of a nucleon accompanied by pions, and nucleon-antinucleon pairs. The first cut lies at $s_{\pi p}=(M+m_{\pi})^2$. The left cut lies at $s_2<-(3M_N)^2$ and corresponds to a $p\bar p p $ state in $u$-channel. Its contribution, for the case $Q^2\sim s\gg M^2$ is suppressed by powers of $(M^2/Q^2)^n$, compared with contributions corresponding to singularities of the right cut, since the operator of higher twist become relevant \cite{LL}. Neglecting the contribution of the left cut the $s_p$ integration contour can be closed to the pole and the $s_p>s_{\pi p}$ cuts, insuring their mutual cancellation. Similar considerations were developed in frame of QED \cite{SUMR}. The application of this result to forward elastic $ep$ scattering amplitude in the high energy limit allows to derive sum rules for the strong interaction contribution to the proton FF. The contribution of the left cut, in this case, has to be precisely calculated. This will be published elsewhere.

Based on this hypothesis (which was proved rigorously in frame of QED and here taken as an assumption) we consider only the QED part of the amplitude corresponding to one nucleon (non-excited) state in the hadronic block and omit the pure strong interaction contributions.

Proton FFs enter in the box amplitude in a form which can be schematically written as
\ba
\int \frac{d^4 k}{i\pi^2 (e)(p)} \frac{F_{1Q}(k^2)+F_s(k^2)}{(k)}\,\,\frac{F_{1Q}(\bar{k}^2)+F_s(\bar{k}^2)}{(\bar{k})},
\ea
$$
\quad
(k)=k^2-\lambda^2, \quad (\bar{k})=(k-q)^2-\lambda^2,
\quad
(e)=(k-p_1)^2-m_e^2, \quad (p)=(k+p)^2-M^2,
$$
where we extract the 'QED' part and do not distinguish Dirac and Pauli form
factors. This integrand expression can be rearranged as
\be
\frac{F_{1Q}(k^2)F_{1Q}(\bar{k}^2)}{(k)(\bar{k})}+\frac{F_s(k^2)F_s(\bar{k}^2)}{(k)(\bar{k})}+
\frac{F_{1Q}(k^2)F_s(\bar{k}^2)}{(k)(\bar{k})}+\frac{F_s(k^2)F_{1Q}(\bar{k}^2)}{(k)(\bar{k})}.
\label{eq:eq4}
\ee
The first term cancels due to the assumption about the 'QED' contribution to the Dirac form-factor (\ref{eq:eqsu}). The second one
can be omitted due to the hypothesis of cancellation in strong interaction. The last two terms coincide and can be written as:
\be
2\frac{F_s(q^2)}{q^2}\int\frac{d^4 k}{i\pi^2 (e)(p)}\frac{1}{(k)}\theta(m_0^2-|k^2|).
\label{eq:eq5}
\ee
The relevant integral is calculated in the Appendix.

The virtual photon emission contribution to the cross section has a form (according to Eq. (\ref{eq:eq4})):
\begin{gather}
\frac{\dd \sigma_v}{\dd\Omega}=
=\frac{\alpha^3}{2\pi t^2M^2\rho^2}a,
\end{gather}
with:
\begin{gather}
a=\int\frac{\dd^4 k}{i\pi^2}\frac{1}{(k)}
\frac{S_T}{(p)}\biggl[
\frac{S_e}{(e)}+\frac{S_{\bar{e}}}{(\bar{e})}\biggr] \theta(m_0^2-|k^2|),
\end{gather}
and
\ba
(\bar{e})&=&(k+p_1')^2-m_e^2,\nonumber \\
S_e&=&\frac{1}{4}Tr\hat{p}_1'\gamma_\mu(\hat{p}_1-\hat{k})\gamma_\nu\hat{p}_1\gamma_\eta,
\nonumber \\
S_{\bar{e}}&=&\frac{1}{4}Tr\hat{p}_1'\gamma_\nu(\hat{p}_1'+\hat{k})\gamma_\mu\hat{p}_1\gamma_\eta,
\nonumber \\
S_T&=&\frac{1}{4}Tr (\hat{p}+M)\Gamma_{\eta}(-q)(\hat{p}'+M)
\Gamma_{\nu}(k) (\hat{p}+\hat{k}+M) \Gamma_{\mu}(q-k),\nonumber \\
\Gamma_{\mu}(q)&=&\left [F_{1}(q^2)+\frac{\hat q}{2M}F_{2}(q^2)\right ]\gamma_\mu.
\ea
Using the formulas given in Appendix, the virtual photon contribution to the differential cross section with two photon exchange, $d\sigma_v$, can be written as:
\be
d\sigma_v^{odd}= \displaystyle\frac{2\alpha}{\pi}\ln\rho\left ( \ln\frac{2 EM}{\lambda^2}
-\displaystyle\frac{1}{2}\ln\rho \right )d\sigma_{Born}.
\label{eq:eqv}
\ee
The IR divergence from virtual photon emission contribution is, as
usually, canceled when summing the contribution from emission of soft real photons:
\begin{gather}
\frac{\dd\sigma^{soft}}{\dd\Omega}=
\biggl[\frac{\dd\sigma_{Bt}}{\dd\Omega}
+\frac{\dd\sigma_{Bbox}}{\dd\Omega}
\biggr]\delta_{soft}^{odd}
=\frac{\dd\sigma_{Bt}^{soft}}{\dd\Omega}
+\frac{\dd\sigma_{Bbox}^{soft}}{\dd\Omega}.
\end{gather}
The quantity $\delta_{soft}^{odd}$
was considered in \cite{china,Ma00}:
\ba
\delta_{soft}^{odd}&=&-2\left . \frac{4\pi\alpha}{16\pi^3}\int\frac{d^3k}{\omega}(\frac{p_1'}{p_1'k}-\frac{p_1}{p_1k})
(\frac{p'}{p'k}-\frac{p}{pk})\right |_{S_0,\omega\le \Delta E}
\nonumber \\
&=&\frac{2\alpha}{\pi}\biggl[\ln\frac{1}{\rho}\ln\frac{2\rho\Delta E}{\lambda}
+\ln x\ln\rho+\Li{2}{1-\frac{1}{\rho x}}-\Li{2}{1-\frac{\rho}{x}}\biggr],
\nonumber \\
x&=&\frac{\sqrt{1+\tau}+\sqrt{\tau}}{\sqrt{1+\tau}-\sqrt{\tau}},
\ea
with $\lambda$, and $\Delta E$ respectively the mass and the maximal energy of the soft photon.

\section{Results and Discussion}

\begin{figure}
\includegraphics[scale=.6]{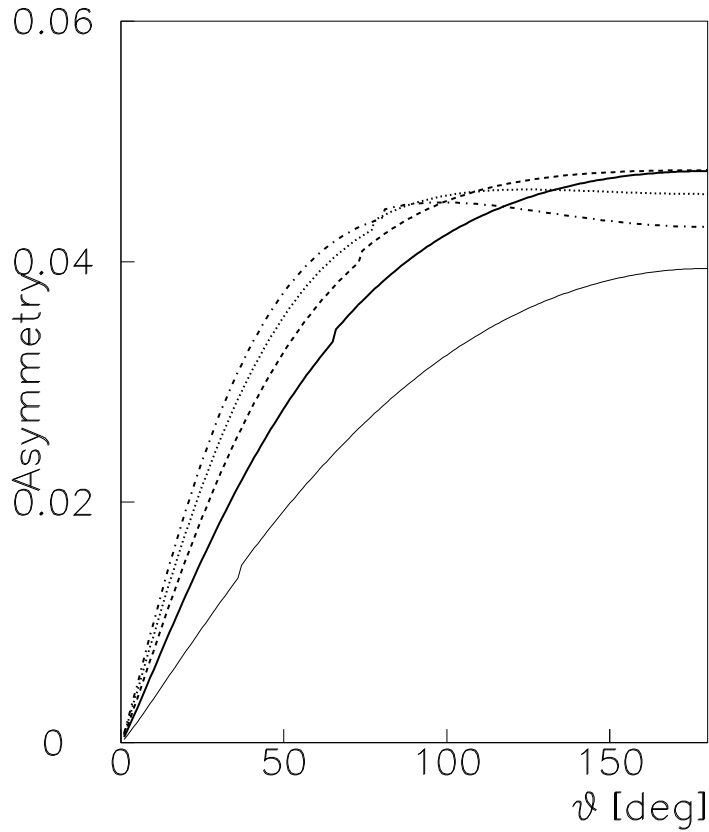}
\caption{Charge odd correlation in electron-positron scattering as a function of the scattering angle $\theta$, for $Q^2=$1 GeV$^2$ (thin solid line), 3 GeV$^2$ (thick solid line), 5 GeV$^2$ (dashed line), 7 GeV$^2$ (dotted line), 9 GeV$^2$ (dash-dotted line). The calculation corresponds to $\Delta E$=0.01 $E$ (see Eq. \protect\ref{eq:eqv1}). }
\label{fig:asym}
\end{figure}

The final result for the asymmetry $A^{odd}$ (\ref{eq:asym}) is:
\ba
A^{odd} &=&\frac{2\alpha}{\pi}
\biggl[\ln\frac{1}{\rho}\ln\frac{(2\Delta E)^2}{ME}-
\frac{5}{2}\ln^2\rho+\ln x\ln\rho+\Li{2}{1-\frac{1}{\rho x}}-\Li{2}{1-\frac{\rho}{x}}\biggr], \nonumber \\
\rho &=&\left (1-\frac {Q^2}{s}\right)^{-1}.
\label{eq:eqv1}
\ea
The finite part of the asymmetry is calculated for different values of $Q^2$ and $\theta$. The results are plotted in Fig. \ref{fig:asym}, as a function of $\theta$. The asymmetry vanishes for $\theta=0$ and reaches the largest values for $\theta=\pi$. It is measurable, of the order of 5 \%. The largest contribution to the asymmetry is given by the first term in Eq. (\ref{eq:eqv1}), which depends on the soft photon energy. The calculation is done for $\Delta E=0.01E$.

\begin{figure}
\includegraphics[scale=.6]{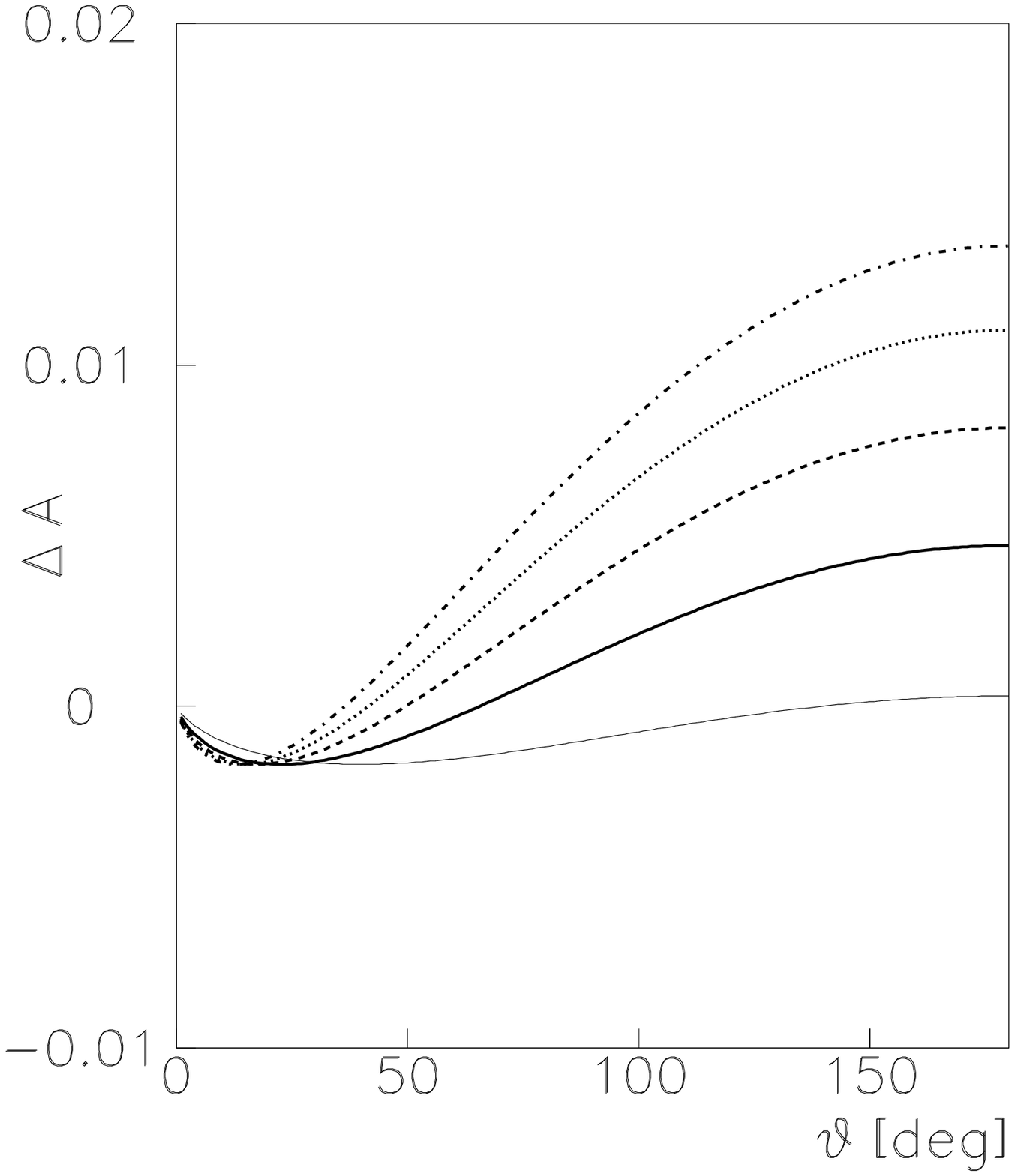}
\caption{ Difference between the present asymmetry and the calculation \protect\cite{Ma00}, $\Delta A$, Eq. \protect\ref{eq:diff}. Notations as in Fig. \protect\ref{fig:asym}.}
\label{fig:diff}
\end{figure}

In Ref. \cite{Ma00}, the approach used to calculate TPE has to be considered as a model: one of the exchanged photon is quasi real. The contribution of inelastic intermediate states is also ignored. The comparison with the results from Ref. \cite{Ma00}, $ A^{odd}_{MT}$,
is shown in Fig. \ref{fig:diff}, where the absolute difference
\be
\Delta A =A^{odd}_{MT}-A^{odd}=\frac{2\alpha}{\pi}\left (\ln\frac{1}{\rho}\ln\frac{s}{Q^2}+\frac{1}{2}\ln^2\rho \right )
\label{eq:diff}
\ee
between the two calculation is shown. The present results are in general smaller, except at small angles, where they are comparable. The main difference is due to the term related to $\ln(Q^2/s)$ which gives a different $\Delta E $ dependent contribution.

\section{Conclusions}

.

Charge asymmetry in electron proton elastic scattering
contains essential information on the contribution of the real part of the $2\gamma$ exchange to the reaction amplitude. This amplitude can shed light on Compton
scattering of virtual photons on proton. It contains a part
corresponding to proton intermediate state, which carries the
information on proton FFs. Another term corresponds to excited nucleon
states and inelastic states such as $N\pi$, $N2\pi$, $N\bar N N$. Their
theoretical investigation is strongly model dependent.

Similar effects of charge and angular asymmetries  can also be due to Z-boson exchange, but such contribution
is small for moderate-high energy colliders. The ratio of corresponding contributions  can be
evaluated as: $\sim (\pi g_Vg_A s)/(\alpha M_Z^2)< 5 \cdot 10^{-3}$ for $s<10$ GeV$^2$ ($g_V$ ($g_A$) is
the vector (axial) coupling constant of the Z boson to fermion).

The analytical calculation of $2\gamma$ amplitude with FFs encounters mathematical difficulties. In Ref. \cite{Bo06}
the results for the box amplitude with arbitrary FFs was investigated.  Other works use different approaches to include FFs, and the results are quantitatively  different \cite{twogamma,Ma00}.

The analytical model presented in this paper is based on two main assumptions:
- the separation of the nucleon FFs in two terms, one of which corresponds to the QED contribution, which contributes explicitly to a narrow region of transferred momenta, close to zero. - the mutual compensation of the strong interaction contributions to the TPE amplitude, arising from nucleon form factors and from inelastic states. The last assumption, which has been proved in QED and which holds for zero scattering angle amplitude, has to be considered as an approximation when applied to the high energy limit of large angle scattering. The tendency of cancellation of elastic proton state and the $\Delta$ resonance in the real part of the TPE amplitude was previously noted in literature (\cite{By07,Ko05}).

The numerical results obtained here are in agreement with our previous calculation \cite{china,By07} and confirm our previous conclusion that two photon contribution can not be responsible for the discrepancy in recent FFs measurements. A more plausible explanation is that higher order radiative corrections should be taken into account in the leptonic vertex \cite{By07}.

Other works \cite{Ko05} devote attention to the excited intermediate states as $\Delta$ and $N^*$ resonances,
introducing additional uncertainties. In our approach, excited states should not be included, as they correspond
to poles in the second physical sheet. Only contribute $n\pi N$ states, with any number of pions.

Our main assumption about the compensation of pure strong interaction induced
contributions to FFs and inelastic channels allows us to avoid additional uncertainty
connected with inelastic channels. Experiments measuring charge-odd observables in $ep$ scattering  will be critical for the verification of the validity of our model.

The numerical results show that charge-odd correlations are of the order of few percent, in the kinematical region considered
here. Such value is expected to be larger at larger $Q^2$ values and could be measured in very precise experiments,
at present facilities.

\section{Acknowledgments}
Two of us (E.A.K. and V.V.B.) acknowledge the kind hospitality of Saclay,
where part of this work was done. This work was partly supported by
 INTAS grant 05-1000-008-8323 and grant MK--2952.2006.2 .

\section{Appendix}
\label{app:A}
The calculation of the integral
\ba
I=\int\frac{\dd^4k}{i\pi^2}\frac{1}{(k)(e)(p)}\theta(m_0^2-|k^2|)
\ea
is performed using the Sudakov's momentum parametrization. For this aim we build two (almost) light-like vectors
\ba
\tilde{p}_1&=&p_1-p\frac{m^2}{s},\tilde{p}=p-p_1\frac{M^2}{s},s=2pp_1, \tilde{p}_1^2=0;\tilde{p}^2=0, \nn \\
2\tilde{p}p&=&M^2;2\tilde{p}_1p_1=m^2.
\ea
The integration is done over the momentum variables:
$$
k=\alpha\tilde{p}+\beta\tilde{p}_1+k_\bot, k_\bot p=k_\bot p_1=0, k^2=s\alpha\beta-\vec{k}^2,
$$
and the phase volume is :
$$
d^4k=\frac{s}{2}d\alpha d\beta d^2\vec{k}.
$$
The analysis of the location of the poles of the denominators in $\alpha$ and $\beta$ planes:
\ba
(k)&=&k^2-\lambda^2=s\alpha\beta-\vec{k}^2-\lambda^2+i0; \nn \\
(e)&=&k-p_1)^2-m^2+i0= s\alpha\beta-\vec{k}^2-s\alpha-m^2\beta+i0; \nn \\
(p)&=&(k+p)^2-M^2+i0=s\alpha\beta-\vec{k}^2+s\beta+M^2\alpha+i0,
\ea
leads to two regions of non zero contribution $0<\alpha,\beta<1$;$-1<\alpha,\beta<0$, with equal contributions.
We obtain
\ba
I=2\int\limits_0^1 d\alpha \int\limits_0^1d\beta\frac{d\vec{k}^2}{i\pi}\frac{1}{s\alpha\beta-\vec{k}^2-\lambda^2+i0}
\frac{1}{-s\alpha-m^2\beta}\frac{1}{s\beta+M^2\alpha}\theta(m_0^2-|k^2|).
\ea
the real part of $I$ (which is relevant for us) can be extracted using the identity $1/(x+i0)={\cal P}(1/x)-i\pi\delta(x)$,
and performing the integration on $\vec{k}^2$:
$$\int \limits_0^{m_0^2} dz \delta(z-a)=\theta(a), a=s\alpha\beta-\lambda^2,m_0^2>a>0. $$
The final answer is
\ba
Re I=2\left [\frac{1}{2}\ln^2\frac{s}{mM}+\ln\frac{s}{mM}\ln\frac{mM}{\lambda^2}-\frac{1}{2}\ln^2\frac{M}{m}-\frac{\pi^2}{2}\right ].
\ea
It is interesting to compare this result with the same integral without cut on $|k^2|$:
\ba
J=Re\int\frac{\dd^4k}{i\pi^2}\frac{1}{(k)(e)(p)}=\ln^2\frac{s}{mM}-\ln^2\frac{M}{m}+2\ln\frac{s}{mM}\ln\frac{mM}{\lambda^2}-\frac{4\pi^2}{3}.
\ea
Using similar expressions for another integral of this type, it is straightforward to recover Eq. (\ref{eq:eqv1}).

\end{document}